# The 120Gbps VCSEL Array Based Optical Transmitter (ATx) Development for the High-Luminosity LHC (HL-LHC) Experiments

**Di Guo**[a,b]**, Chonghan Liu**[b]**, Jinghong Chen**[c]**, John Chramowicz**[d]**, Binwei Deng**[b,e]**, Datao Gong**[b]**, Suen Hou**[f]**, Ge Jin**[a]**, Simon Kwan**[d]**, Futian Liang**[a]**, Xiaoting Li**[b,g]**, Gang Liu**[b,h]**, Tiankuan Liu**[b]**, Alan Prosser**[d]**, Da-Shung Su**[f]**, Ping-Kun Teng**[f]**, Tongye Xu**[i]**, Jingbo Ye**[b]**, Xiandong Zhao**[b]**, Annie C. Xiang** [b,*] **and Hao Liang** [a,*]

[a] *State Key Laboratory of Particle Detection and Electronics,*
  *University of Science and Technology of China,*
  *Hefei Anhui 230026, China*

[b] *Department of Physics, Southern Methodist University,*
  *Dallas, TX 75275, USA*

[c] *Department of Electrical Engineering, Southern Methodist University,*
  *Dallas, TX 75275, USA*

[d] *Electronic Systems Engineering Department, Fermilab,*
  *Batavia, IL 60510, USA*

[e] *Hubei Polytechnic University,*
  *Huangshi, Hubei 435003, P. R. China*

[f] *Institute of Physics, Academia Sinica,*
  *Nangang 11529, Taipei, Taiwan*

[g] *Department of Physics, Central China Normal University,*
  *Wuhan, Hubei 430079, P.R. China*

[h] *Institute of High Energy Physics, Chinese Academy of Sciences,*
  *Beijing 100049, P. R. China*

[i] *MOE Key Lab on Particle Physics and Particle Irradiation, Shandong University,*
  *Ji'nan 250100, China*
  *E-mail:* cxiang@smu.edu and simonlh@ustc.edu.cn

ABSTRACT: The integration of a Verticle Cavity Surface-Emitting Laser (VCSEL) array and a driving Application-Specific Integrated Circuit (ASIC) in a custom optical array transmitter module (ATx) for operation in the detector front-end is constructed, assembled and tested. The ATx provides 12 parallel channels with each channel operating at 10 Gbps. The optical transmitter eye diagram passes the eye mask and the bit-error rate (BER) less than $10^{-12}$ transmission is achieved at 10 Gbps/ch. The overall insertion loss including the radiation induced attenuation is sufficiently low to meet the proposed link budget requirement.

---

[*] Corresponding authors.

**Contents**

## 1. Introduction

Collision rates available at high-energy physics experiments have increased considerably with the turn on of the Large Hadron Collider (LHC). To process the resulting data in real time, highly complex algorithms have been applied. This trend is foreseen to continue with the high luminosity LHC (HL-LHC) developments. As a result, data rates proposed for collider detector readout upgrades have increased rapidly. Fibre optical links became prevalent and several commercial parallel optical interconnect modules have been investigated for use in the off-detector electronic systems. The main drives for the adoption of parallel optics are high bandwidth, high channel density and low power consumption, whereas the main challenges are radiation and electromagnetic tolerances and system qualification.

    The harsh radiation environment in which the front-end optical link modules would be operated demands innovative customization at the sub-assembly level and rigorous qualification at the link level. Most experiments have higher bandwidth requirements in uplink (from detector to the counting room) than in downlink (to the detector). Therefore there is a greater need for the development of parallel transmitters than for parallel receivers. Industry employs two mainstream platforms for parallel transmitters: (1) ones that make a vertical to lateral turn in the optical path as exemplified in the MicroPOD of Avago [1], and (2) ones that make the 90 degree turn in the electrical path [2] as exemplified in some commercial QSFP [3] and SNAP12 [4] form factor transmitters. The former structure has been the subject of this project and the latter structure has been investigated by other groups in the community [5]. Radiation tolerance has been addressed at the subcomponent level. Both surface emitting and edge emitting lasers made from GaAs quantum wells demonstrate tolerance to Non-Ionizing Energy Loss (NIEL) up to $10^{15}$ cm$^{-2}$ 1-MeV neutrons equivalents, with increased threshold, decreased slope efficiency, but comparable frequency response [6]. Other passive materials suffer radiation induced

attenuation (RIA) that can be quantified by a conservative figure of merit [7]. Radiation-tolerant ASICs have been custom designed in CMOS [8] or SOS technologies [9] up to 8 Gbps whereas 10 Gbps per channel array drivers are under developments.

In this paper we present the development of the ATx module, an array based parallel optical transmitter that is composed of a 12-channel VCSEL array, an array driver ASIC, a carrier substrate and a micro-lens array with matching ribbon cable. The adaptation and analysis of a mass produced Module Optical Interface (MOI) and a Prizm fibre were described in Section **2**. To achieve high-speed operation up to 10 Gbps per channel, FR4 and ceramic substrates with high-density transmission lines were designed as discussed in Section **3**. The manufacturing assembly process employing a precision pick and place is examined in Section **4**. To evaluate the DC and AC characteristics of the complete module, we measured optical eye diagrams and bit-error rate (BER) curves as received by a 10-Gbps receiver. Results are presented in Section **5**, along with the X-ray irradiation results. The Power budget for a prospective ATx to commercial receiver link is discussed in Section **6**. Conclusions are presented in Section **7**.

## 2. Optical interface

The optical interface is composed of an MOI and a Prizm fibre, as shown in Figure 1(a) for a cross sectional view. The prizm fibre is a ribbon fibre with one end terminated with the Prizm connector, and the other end terminated with the Multiple Termination Push-on (MTP) connector. Beams emitted from the VCSEL array are collimated by 12 micro lenses of the MOI and deflected 90 degrees by the total internal reflection (TIR) microprizms within the Prizm connector, which can be clipped onto the MOI. Optical path alignment between the Prizm connector and the MOI is guaranteed by guide pin holes and posts. PRIZM and LightTurn are trademarks of US Conec Ltd. And both the MOI and the Prizm fibre are low-cost, commercially available standard components.

A prerequisite of the optomechanical design is that the component alignment tolerances have to be achievable by feasible assembly methods. The MOI alignment tolerances on vertical and lateral positioning relatively to the VCSEL sources are ± 10 μm and ± 20 μm respectively. Using a spacer with minimal adhesive thickness, these tolerances can be met by a precision pick and place machine. Figure 1(b) is a photo of an assembled ATx module showing an MOI aligned and tacked down to the substrate. Figure 1(c) shows the complete optical components with the Prizm connector clipped onto the MOI.

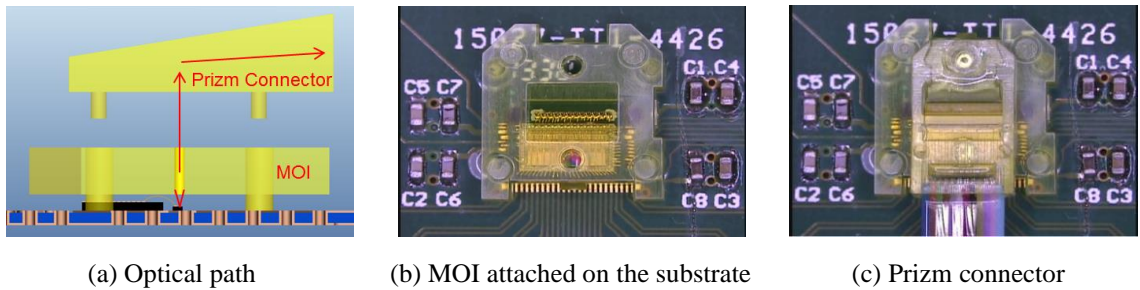

(a) Optical path      (b) MOI attached on the substrate      (c) Prizm connector

**Figure 1**. Optical path and photos of MOI and Prizm connector

# 3. Electrical interface

The substrate (carrying the VCSEL, the driver ASIC, and the MOI) provides 12 pairs of high-speed, high-density differential signals, configuration I/O, and power traces to constitute the electrical interface. Two different versions of electrical interfaces have been designed to offer reflow or pluggable options (an edge wrap version and a mezzanine connector version).

The four-layer FR4 Print Circuit Board (PCB) substrates were fabricated first. The edge wrap FR4 substrate is 1.9 cm × 2.2 cm × 0.9 mm in size and has 52 half vias around four edges serving as the electrical interconnections. All the traces are on the top layer, and a large area of exposed copper is poured on the bottom layer with plated through vias connected to the die-carrying pads on the top layer. The copper will make full contact with the metal pad on the main board as a means of heat dissipation. The 3-D modules in Figures 2(a) and 2(b) show the structure of this substrate. Twelve pairs of high-speed differential transmission lines are implemented with 5 mil wire width and 3 mil wire spacing, and adjacent differential pairs are separated by 16 mil throughout most of their length to minimize crosstalk. Figure 3(a) is a photo showing the edge wrap version of ATx that has already been soldered onto the testing board.

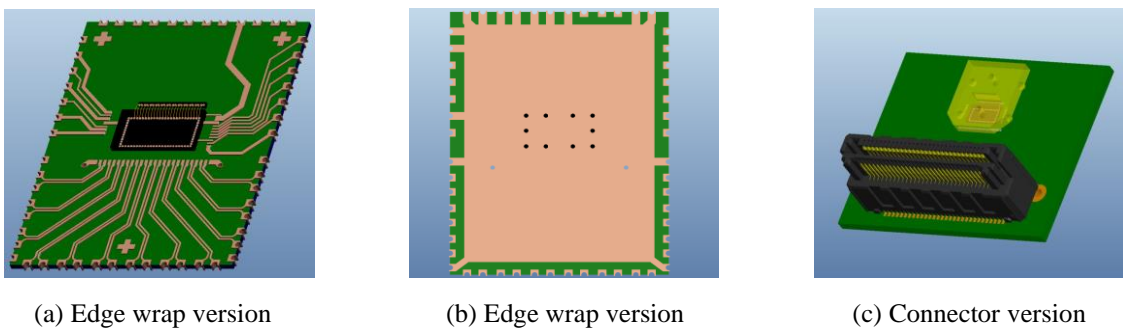

(a) Edge wrap version  (b) Edge wrap version  (c) Connector version

**Figure 2**. 3-D modules of two versions of the ATx substrate

The mezzanine connector version FR4 substrate is 2.5 cm × 2.0 cm × 0.9 mm in size. All the electrical connections are implemented through a high-speed, high-density mezzanine connector, as shown in Figure 2(c). Twelve differential pairs are separated from their nearest neighbors in the upper row of the connector by ground pins. I/O and power are arranged at the lower row pins. Figure 3(b) is a photo of the connector version of ATx.

An edge wrap ceramic substrate is also designed and fabricated using single layer aluminum nitride thick-film technology. Figure 3(c) is a photo of the ceramic substrate before dies are attached and the MOI is assembled. The ceramic substrate supports smaller line width and spacing- we can achieve 3 mil wire width and 2.2 mil wire spacing on the top layer with large area of copper poured on the bottom layer serving as the ground plane to maintain the characteristic impedance for high speed differential pairs. The size of the substrate is 9.8 mm × 13 mm × 0.3 mm, which is smaller and thinner than the former FR4 substrates. In addition, low thermal coefficient of linear expansion of ceramic supports better reliability, and a much flatter surface than an FR4 PCB is also favorable to the MOI attachment which will be described below.

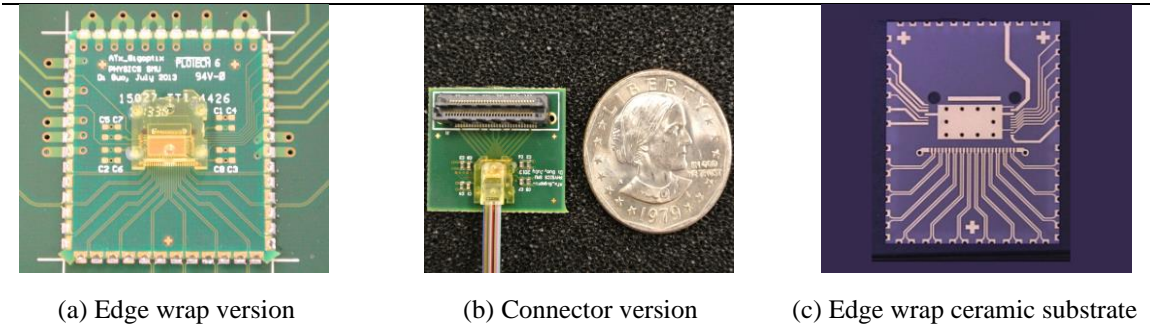

(a) Edge wrap version      (b) Connector version      (c) Edge wrap ceramic substrate

**Figure 3**. Photographs of two versions of the ATx module

## 4. Assembly

The ATx module assembly process consists of die attachment, wire bonding, MOI attachment and reflow soldering. Care is needed for each step to preserve the signal integrity and the optical alignment.

The space between the VCSEL array and the driver die is designed to be as close as 220 μm to shorten the bonding wire to support 10 Gbps transmission. Nonconductive glue is used to attach dies to the substrate to prevent wire bonding pads from shorting. The loop height of bonding wire from the driver to the VCSEL array also needs to be closely controlled, due to the thickness difference of two dies (150 μm and 250 μm) and the restriction from the height of the MOI posts (442 μm), which can be seen from Figure 1(a).

MOI attaching steps, including alignment to the VCSEL array and attachment to the substrate, represent the most crucial procedure for optimal coupling of the optical signals. For the alignment step, it is impossible to directly align the centers of the lenses with the VCSEL light emitting apertures by seeing through the lenses because the lenses are out of focus. To solve this problem, the datum line A (the axis through 12 VCSEL light emitting apertures) is first determined using the microscope reticle, and the datum line B (the axis through the 12 center points of MOI lenses) is then identified after placing the MOI over the VCSEL array. The MOI is adjusted to make the datum line B coincident with the datum A to perform the alignment. Figure 4 shows the two datum lines we need to identify under the microscope.

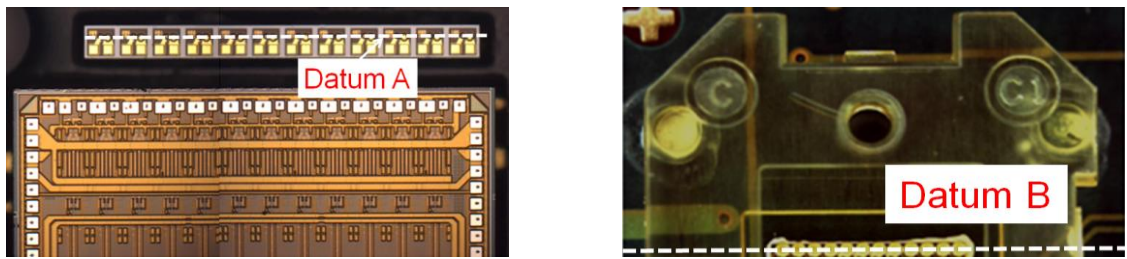

**Figure 4**. Auxiliary datum lines for the MOI alignment

For the MOI attachment step, the epoxy is previously applied to the substrate where the MOI posts will be positioned before the MOI is placed down. An ultraviolet (UV) light and/or thermal curable epoxy (Part No. 3410 from Optocast) is chosen, which permits a quick UV pre-cure to fix the MOI while the MOI is still under the microscope to minimize the potential shifting during the curing. After the MOI is aligned and pre-cured, the whole module will be moved for thermal curing under 125 °C for 20 minutes to improve bonding strength.

A reflow process is necessary when we solder the 60-pin connector in the module of connector version or for mass application of the edge wrap modules. In order to accommodate the reflow procedure in the assembly, low temperature soldering paste (for example Sn42/Bi58) is chosen to achieve the soldering with the highest temperature under 150 °C, which will not cause any functional degradation to the optical components.

## 5. Test Results

Results reported here were tested from the edge wrap FR4 version of ATx. The connector version has been assembled and is under testing, and the ceramic one is to be assembled. The VCSEL driver now used in the ATx is the Gigoptix HXT5112A, while the VCSEL array is the ULM850-10-TT-N0112U.

### 5.1 Optical test

A testing carrier board was designed to provide signals for 6 channels of the ATx, 2 via SMA (Channel 2, 11) and 4 via FPGA Mezzanine Card (FMC). The block diagram and a picture of the optical eye test setup are shown in Figure 5. Eye diagrams of two channels of ATx close to the two ends (Channel 2, 11) were captured by an oscilloscope. Both channels showed widely open eyes for a 10 Gbps Pseudo-Random Binary Sequence $2^7$-1 (PRBS7) pattern as shown in Figure 6.

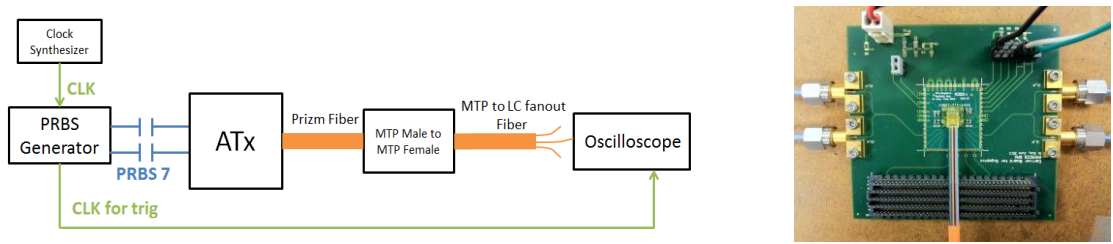

**Figure 5**. Test setup of optical eye-diagram experiment

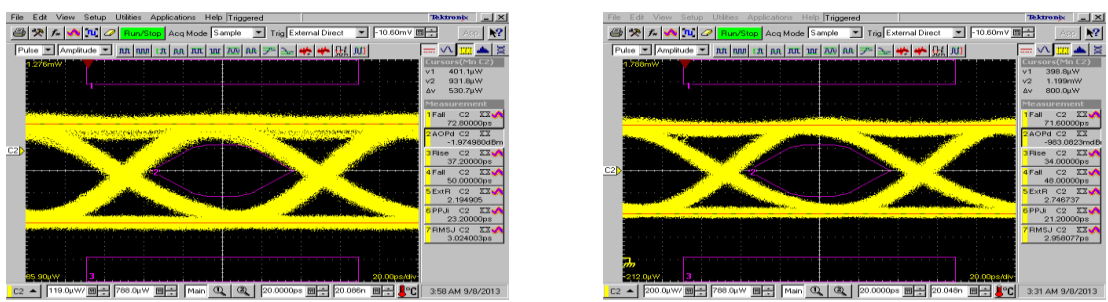

(a) Optical eye of Ch2 at 10Gbps        (b) Optical eye of Ch11 at 10 Gbps

**Figure 6**. Optical eye diagrams of the ATx at 10 Gbps

The link bit error rate (BER) test of ATx (Channel 11) and that of a commercial SFP+ transmitter have been measured and compared. The block diagram and a picture of the test setup are shown in Figure 7. The same SFP+ Rx module was used as the receiver during the test. Figure 8 shows a system penalty of 3 dB at the BER of $10^{-12}$ and 10 Gbps. This is an overall system penalty, which can be influenced by different VCSEL drivers, VCSELs, coupling methods, high speed signal paths and die packages. Note that the transmission lines in the ATx testing were much longer than those in the SFP+ Tx, and an edge wrap interface in the propagation path may also contribute to the penalty.

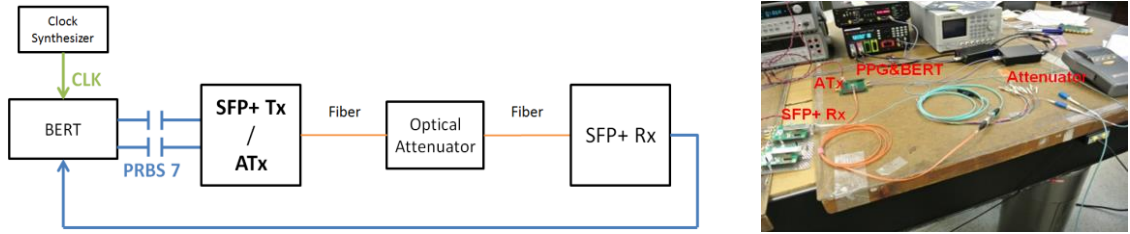

**Figure 7**. Block diagram and picture of the link BER test setup

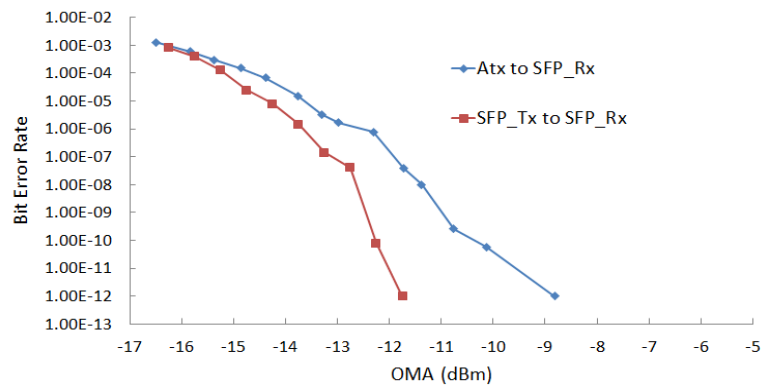

**Figure 8**. Link BER test results at 10 Gbps

Out of the edge wrap and mezzanine connector ATx modules assembled, one MOI was misaligned. Misalignment between the MOI and the VCSEL array caused the 850 nm light to scatter into the body of the Prizm connector instead of coupling into the fibre, and the light was clearly seen from a camera at the joining section of the Prizm connector. One possible reason for the MOI misalignment is the localized flatness variation of the substrate, which impacts both the MOI placement during the alignment and the die attachment procedure. FR4 substrates tend to have this type of surface variations. The consistency of the MOI assembly is expected to be improved in the ceramic substrate version.

## 5.2 Total insertion loss and X-ray test

As previously stated, radiation tolerances of most components of the ATx module have been investigated except the MOI and the Prizm connector. From vendor supplied information, they are made from polyetherimide (PEI), which is sufficiently radiation tolerant in mechanical strength. Optically, both the refractive index change and radiation induced attenuation have been observed in polymers as the result of ion bombardment [10]. Displacement induced effects are only demonstrated at very high threshold. Therefore, an X-ray test was conducted on a fully assembled ATx module. The test setup was similar to that of the optical eye test as shown in Figure 5. A fully assembled ATx module with a complete optical link was prepared, and the transmitted light power was measured before and after the X-ray exposure. The VCSEL array was DC driven throughout the test. Light power-current (L-I) curves of 4 channels were recorded at different total dose levels, and Figure 9 shows the two of them.

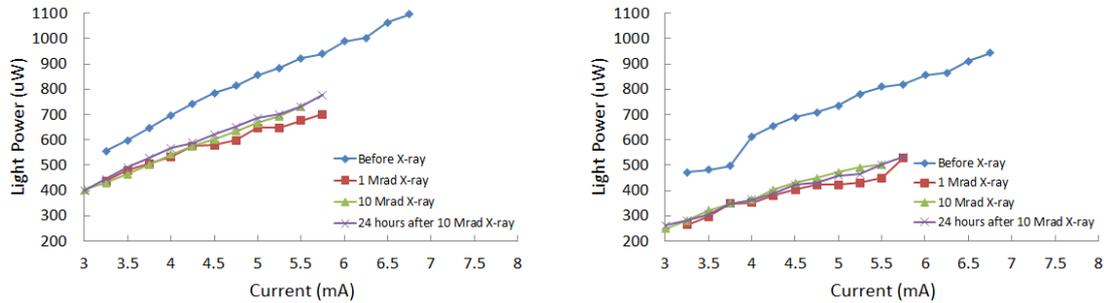

**Figure 9**. L-I characteristic curves

Before the X-ray exposure, when each channel of the ATx was driven by a 6 mA direct current, the obtained light powers of 4 channels were between 1010 μW and 856 μW. According to the VCSEL array specification, a 6 mA driving current corresponds to an output of 2000 μW, indicating the total insertion loss of a fully assembled ATx module is between 3 dB to 3.7 dB.

At a dose rate of 330 rad($SiO_2$)/s, additional attenuation was observed, increasing rapidly to contribute another 1.5 dB at the total dose of 1 Mrad($SiO_2$) where the attenuation saturated. No further attenuation was observed as the total dose was increased up to 10 Mrad($SiO_2$). Twenty-four hours after the radiation, no recovery (reduction in the attenuation after irradiation) was observed. The attenuation observed could consist of contributions of both ionizing degradation (including attenuation and change of reflectivity) and mechanical degradation such as placement shift and coupling precision. Also the test was conducted under a very high dose rate, and future work is needed to carry out the characterization of MOI and Prizm connector ionizing induced attenuation in various radiation environments. The 2 dB degradation is interpreted as a maximum allocation for the assembly radiation degradation.

## 6. System Consideration

The ATx is designed to work with commercial parallel receivers at the back end, which comply with the IEEE 100GBASE-SR10 specifications. As the standard for the 100GBASE-SR10 links is emerging, it is recognized that the links are constrained by both link bandwidth and component jitter [11]. Table 1 lists an estimated power budget allocation for transmission from an ATx to 100GBASE-SR10 receiver. The link penalties are derived from the 10Gbps Ethernet link model, accounting for various effects including intersymbol interference (ISI), mode partition noise, modal noise, and relative intensity noise. For the transmission to stretch from 100 meters to 150 meters over OM3/OM4 fibres, link penalty is increased by another 1 dB. The jitter penalty is estimated by projecting the horizontal eye opening to the vertical eye opening. A 0.24 UI timing variation is equivalent to ~2dB closure, and another ~2dB penalty is assigned to the receiver. Fibre loss is derived from a 3 dB/km maximum attenuation figure of merit. Connector insertion loss is derived from 0.5 dB per breakpoint with maximum of three connections. The receiver sensitivity of the commercial parallel receivers is around -11.3 dB. In order to obtain the 9.3 dB power budget, the transmitter output (output of a fully assembled

ATx module) needs to be over -2 dBm, which can be achieved after considering a maximum 6 dB total loss of the ATx module.

|  | Tx over Rx | Attenuation | Insertion loss | Link penalty | Jitter allocation | Link margin |
|---|---|---|---|---|---|---|
| Budget (dB) | 9.3 | 0.4 | 1.5 | 1.9 | 4.0 | 1.5 |

Table 1. Power budget allocation

## 7. Conclusion

This paper has demonstrated an ATx prototype, the 12-channel parallel optical transmitter module, operating at 10 Gbps and aggregate data rate of 120 Gbps. The optical transmitter eye diagram passes the eye mask without a single violation and the transmitted BER less than $10^{-12}$ is achieved with a commercial receiver through standard 50 μm MMF ribbon with an uncoded PRBS7 pattern. The optomechanical design is a conventional low cost microelectronics assembly method via precision pick and place machine. The total insertion loss is less than 4 dB, and the radiation induced attenuation is less than 2 dB, which is sufficient to meet the proposed link budget requirement. A custom designed driver ASIC will replace the current commercial one soon, and another active alignment method is under development. Statistical and environmental characterization as well as mechanical designs regarding heat dissipation and dust prevention will also be carried out in future development.

**Acknowledgments**

This work is supported by the US Department of Energy Collider Detector Research and Development (CDRD) data link program. The authors also would like to thank Jee Libres and Alvin Goats of VLISP, Phil Melen of DITF and Alan Ugolini of USConec for informative discussions.